\documentclass[conference]{IEEEtran}
\ifCLASSINFOpdf
  \usepackage[pdftex]{graphicx}
\else
  \usepackage[dvips]{graphicx}
  \graphicspath{{../figs/}}
  \DeclareGraphicsExtensions{.eps}
\fi
\begin{document}

%
\title{Spectrum Sensing Based on Blindly Learned Signal Feature}

\author{\IEEEauthorblockN{Peng Zhang, Robert C. Qiu}
\IEEEauthorblockA{Department of Electrical and Computer Engineering\\
Cookeville, TN 38505\\
Tennessee Technological University\\
Email: $\{$pzhang21,rqiu$\}$@tntech.edu}}
\maketitle

\begin{abstract}
Spectrum sensing is the major challenge in the cognitive radio (CR). We propose to learn local feature and use it as the prior knowledge to improve the detection performance. We define the local feature as the leading eigenvector derived from the received signal samples. A feature learning algorithm (FLA) is proposed to learn the feature blindly. Then, with local feature as the prior knowledge, we propose the feature template matching algorithm (FTM) for spectrum sensing. We use the discrete Karhunen--Lo{\`e}ve transform (DKLT) to show that such a feature is robust against noise and has maximum effective signal-to-noise ratio (SNR). Captured real-world data shows that the learned feature is very stable over time. It is almost unchanged in 25 seconds. Then, we test the detection performance of the FTM in very low SNR. Simulation results show that the FTM is about 2 dB better than the blind algorithms, and the FTM does not have the noise uncertainty problem.
\end{abstract}

\section{Introduction}
\label{introduction}

Radio frequency is fully allocated for the primary users (PU), but with low utilization rate \cite{staple2004end} as low as $15\%$ \cite{force2002spectrum, cabric2004implementation}. The concept of cognitive radio (CR) was proposed so that the secondary users (SU) can occupy the unused spectrum from the PU, therefore improving the spectrum utilization rate. The CR requires the SU to detect the existence of PU in a short time in very low signal-to-noise ratio (SNR), which is spectrum sensing. IEEE 802.22 is the first IEEE working group embedding the CR technology \cite{ieee80222} and has triggered lots of research. 

\begin{figure}[tbp]
	\centering
		\includegraphics[width=0.50\textwidth]{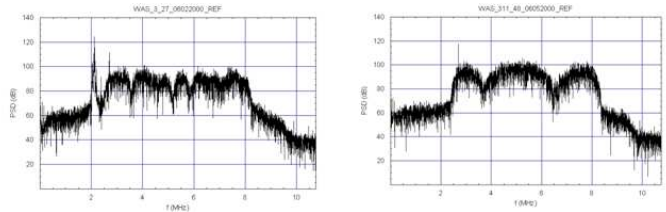}
	\caption{Spectrum measured at different locations in Washington D.C.. Left: `Single Family Home'; Right: `Apartment (High-Rise)'. The pilot tones are located at different frequency locations. Two spectrum suffer different frequency selective fading.}
	\label{fig:Two_Spectrum_WAS}
\end{figure}

Many spectrum sensing algorithms have been proposed. Take the DTV signal sensing for example. Specific features are often defined from the spectral information, such as pilot tone \cite{cordeiro2007spectrum}, spectrum shape \cite{Quan2009ss} and cyclostationarity \cite{dandawate1994statistical}, etc.. Generally speaking, they are robust and have good performance when it is assumed that those features are universal to all the SUs. However, such assumption is not true in real-life. Fig. \ref{fig:Two_Spectrum_WAS} shows the DTV spectrum measured \cite{DTV2006Measurements} at different locations. It can be seen that the spectral features are location dependent, due to different channel characteristics and synchronization mis-match, etc. Therefore, we cannot rely on the pre-determined prior knowledge of signals for spectrum sensing. 

Energy based algorithms do not have such problem. Essentially, energy based algorithms require the prior knowledge of noise. However, the noise uncertainty problem \cite{tandra2005fundamental} will limit the performance of energy based algorithms. Pure blind algorithms have been proposed in \cite{zeng2007maximum,lim2008glrt}, such as the maximum eigenvalue to minimum eigenvalue ratio (MME). No noise information is required and the noise uncertainty problem is successfully avoided. 

In this paper, we propose to use learned prior knowledge to improve detection performance. The learned prior knowledge is the leading eigenvector derived from the received signal's sample covariance matrix, using the discrete Karhunen--Lo{\`e}ve transform (DKLT). Similar to the terminology of the pattern recognition in machine learning, we define the leading eigenvector as signal feature. We first propose a feature learning algorithm (FLA) to acquire the local feature blindly. Then, we propose a feature template matching algorithm (FTM) which uses the learned feature for spectrum sensing. The leading eigenvector, a.k.a., feature, is optimum in signal representation \cite{watanabe1965kle} and most reliable when the distribution of signal is unknown \cite{young1971reliability}. In analogy with the recognition of pattern features in image and speech, etc., spectrum sensing is the recognition of the PU feature at the receiver. We will show that:
\begin{enumerate}
	\item Feature is a robust approximation of the non-white wide-sense stationary (WSS) signal against the white Gaussian noise (WGN). 
	\item Feature has maximum effective SNR.
	\item The proposed algorithms are immune to the noise uncertainty problem.
\end{enumerate}
We use both simulated data and real-world data to demonstrate that feature is robust and stable against noise and feature can be learned blindly even in very low SNR. DTV samples \cite{DTV2006Measurements} are used to compare the detection performance of the FTM and the MME. The simulation results show that to achieve the same detection performance, the minimum required SNR for the FTM is about 2 dB lower than that of the MME, which shows that with the feature as the prior knowledge, the detection performance can be improved.


The paper is organized as follows. The FLA and the FTM are presented in Section II. The theoretical background of the feature is introduced in Section III. Simulation results are shown in Section IV and conclusions are made in Section V.

\section{Problem Formulation and the Proposed Algorithms}
The spectrum sensing problem can be modeled as follows. $x\left( t \right) = s\left( t \right) + n\left( t \right)$ represents the received signal at the SU, with $s\left( t \right)$ the PU signal and $n\left( t \right)$ the WGN. Both $s\left( t \right)$ and $n\left( t \right)$ are independent random process with zero mean, but $s\left( t \right)$ is non-white WSS while $n\left(t \right)$ is white Gaussian. Assume the frequency bandwidth being sensed is $B$ centered at frequency $f_c$. After Nyquist sampling with period $T_s  \le 1/B$, we can represent the received signal $x\left(t \right)$ in discrete form: $x\left[ n \right] = x\left( {nT_s } \right)$, $s\left[ n \right] = s\left( {nT_s } \right)$ and $w\left[ n \right] = w\left( {nT_s } \right)$. The spectrum sensing problem has two hypotheses: ${\bf H}_0$, signal does not exist; and ${\bf H}_1$, signal exists. The received discrete form signal under the two hypotheses is therefore as follows:
\begin{equation}
\label{hyp1}
 {\bf H}_0 :{\rm    }x\left[ n \right] = w\left[ n \right]
\end{equation}

\begin{equation}
\label{hyp2}
 {\bf H}_1 :{\rm    }x\left[ n \right] = s\left[ n \right] + w\left[ n \right]
\end{equation}

Two probabilities are of interest. The detection probability, $P_d \left( {{\bf H}_1 |x \left[ n \right] = s\left[ n \right] + w\left[ n \right]} \right)$, and the false alarm probability, $P_f \left( {{\bf H}_1 |x \left[ n \right] = w\left[ n \right]} \right)$.

It is assumed that the learning and sensing processes are performed within the channel coherent time, and there is a pre-whitening filter before any processing.

Let ${\bf x}_n$, ${\bf s}_n$ and ${\bf w}_n$ be random vectors consisting of $N$ samples of $x\left[ n \right]$, $s\left[ n \right]$ and $w\left[ n \right]$, respectively:
\begin{equation}
\label{vector_x}
{\bf x}_n  = \left[ {x\left[ n \right],x\left[ {n + 1} \right], \cdots ,x\left[ {n + N - 1} \right]} \right]^T
\end{equation}

\begin{equation}
\label{vector_s}
{\bf s}_n  = \left[ {s\left[ n \right],s\left[ {n + 1} \right], \cdots ,s\left[ {n + N - 1} \right]} \right]^T
\end{equation}

\begin{equation}
\label{vector_w}
{\bf w}_n  = \left[ {w\left[ n \right],w\left[ {n + 1} \right], \cdots ,w\left[ {n + N - 1} \right]} \right]^T
\end{equation}
where $\left(  \cdot  \right)^T $ denotes matrix transpose. Using $E\left[  \cdot  \right]$ as the notation for expectation, we have corresponding covariance matrices: 

\begin{equation}
\label{Rx}
{\bf R}_{\bf x}  = E\left[ {{\bf x} _n {\bf x} _n^T } \right]
\end{equation}

\begin{equation}
\label{Rs}
{\bf R}_{\bf s}  = E\left[ { {{\bf s} _n {\bf s} _n ^T } } \right]
\end{equation}

\begin{equation}
\label{Rw}
{\bf R}_{\bf w}  = E\left[ {{{\bf w} _n {\bf w} _n^T } } \right]
\end{equation}

Since $s \left[ n \right]$ and $w \left[ n \right]$ are independent, we have:

\begin{equation}
\label{RxRsRw}
{\bf R}_x  = {\bf R}_s  + {\bf R}_w 
\end{equation}

Since $w \left[ n \right]$ is WGN, we have:
\begin{equation}
\label{RI}
{\bf R}_w  = \sigma _w^2 {\bf I}_N 
\end{equation}
where $\sigma _w^2$ is noise variance and ${\bf I}_N$ is $N \times N$ identity matrix.

If we do the eigen-decomposition on the covariance matrix ${\bf R}_x$, we can get a set of eigenvalues $\left\{ {\lambda _1 ,\lambda _2 , \cdots ,\lambda _N } \right\}$ and eigenvectors $\left\{ {{\bf \phi }_1 ,{\bf \phi }_2 , \cdots ,{\bf \phi }_N } \right\}$, satisfying:
\begin{equation}
\label{lambda_set}
\lambda _1  \ge \lambda _2  \ge  \cdots  \ge \lambda _N 
\end{equation}
and

\begin{equation}
\label{eigen_equation}
{\bf R}_x {\bf \phi }_i  = \lambda _i {\bf \phi }_i \begin{array}{*{20}c}
   {} & {}  \\
\end{array}i = 1,2, \cdots ,N
\end{equation}

In the terminology of pattern recognition, $\left\{{\bf \phi } _i\right\}$ are called features. The process of calculating features is called \textsl{feature extraction}. Since our algorithms only deal with the leading eigenvector, in this paper only ${\bf \phi} _1$ is named as feature for brevity.

The exact covariance matrix ${\bf R}_x $ cannot be derived in practice because we do not know the exact expectation of all the random processes. Alternatively, if we define ${\bf \chi }_m = \left\{ {{\bf x}_{1 + N_s  \times \left( {m - 1} \right)} ,{\bf x}_{2 + N_s  \times \left( {m - 1} \right)} , \cdots ,{\bf x}_{N_s  + N_s  \times \left( {m - 1} \right)} } \right\}$ as the $m$-th sensing segment, we can have an approximated sample covariance matrix ${\bf \tilde R}_{x,m}$ by averaging:

\begin{equation}
\label{Rx_segment}
{\bf \tilde R}_{x,m}  = \frac{1}{{N_s }}\sum\limits_{i = 1 + N_s  \times \left( {m - 1} \right)}^{i = N_s  + N_s  \times \left( {m - 1} \right)} {{\bf x}_i {\bf x}_i ^T } 
\end{equation}

Now we use $\varphi _m$ to represent the leading eigenvector of covariance matrix ${\bf \tilde R}_{x,m}$ of segment $m$, a.k.a., the feature of segment $m$. We use the intuitive template matching to find the similarity of features between segments $i$ and $j$: 
\begin{equation}
\label{similarity}
\rho _{i,j}  = \mathop {\max }\limits_{l = 1,2,...,N - k + 1} | {\sum\limits_{k = 1}^N {\varphi _i \left[ k \right]\varphi _j \left[ {k + l} \right]} } |
\end{equation}

Based on the above notations and concept of feature, we propose the FLA and the FTM:

\paragraph{Algorithm 1, the FLA}

\begin{enumerate}
	\item Collect two consecutive sensing segments ${\bf \chi }_i$, ${\bf \chi }_{i + 1}$, with $N_s + N - 1$ samples each.
	\item Compute covariance matrices for each segment according to (\ref{vector_x}) and (\ref{Rx_segment}).
	\item Extract features $\varphi _i$ and $\varphi _{i +1}$ for the corresponding segments.
	\item Compute similarity $\rho _{i,i + 1}$ between these two features using (\ref{similarity}).
	\item If $\rho _{i,i + 1}  > T_e$, then feature is learned as $\varphi _{feature} = \varphi _{i + 1}$.
\end{enumerate}
where $T_e$ is the threshold determined according by the similarity of consecutive noise segments.

Using the learned feature $\varphi _{feature}$ as the prior knowledge, we propose the FTM:

\paragraph{Algorithm 2, the FTM}
\begin{enumerate}
	\item Collect $N_s + N - 1$ consecutive samples.
	\item Compute covariance matrix for this segment according to (\ref{vector_x}) and (\ref{Rx_segment}).
	\item Extract feature $\varphi _{current}$ of the current segment.
	\item Compute similarity $\rho _{feature, current}$.
	\item If $\rho _{feature, current} > T _f$, then hypothesis $\bf H_1$ is claimed. Otherwise, hypothesis $\bf H_0$ is claimed.
\end{enumerate}
Both $T_e$ and $T_f$ are thresholds to be set according to noise statistics. As will be shown later, $T_e$ and $T_f$ are independent to the noise energy, or the SNR. There is no noise uncertainty problems in setting $T_e$ or $T_f$. 

So far we have proposed the FLA and FTM. In the next section we will show why we use leading eigenvector as feature. 

\section{Theoretical Background}
The theoretical background of the FLA and FTM lies in DKLT. It explains why we define feature as leading eigenvector. We follow the description of \cite{therrien1992discrete} to get a brief review of DKLT. If we consider a zero mean random sequence $\left\{ {x\left[ n \right];n = 1, \cdots ,N} \right\}$, this sequence can be expanded in any set of orthonormal basis functions 
$f _i \left[ n \right]$ as:

\begin{equation}
\label{x_expansion}
x\left[ n \right] = \kappa _1 f _1 \left[ n \right] + \kappa _2 f _2 \left[ n \right] +  \cdots  + \kappa _N f _N \left[ n \right]
\end{equation}
where the $\kappa _i $ are coefficients in the expansion and ``orthonormal'' means that the functions satisfy the relation

\begin{equation}
\label{orthonormal}
\sum\limits_{n = 1}^N {f_i^* \left[ n \right]f_j \left[ n \right]}  = \left\{ \begin{array}{l}
 1\begin{array}{*{20}c}
   {} & {}  \\
\end{array}i = j \\ 
 0\begin{array}{*{20}c}
   {} & {}  \\
\end{array}i \ne j \\ 
 \end{array} \right.
\end{equation}
where $*$ denotes conjugate. Following (\ref{x_expansion}) and (\ref{orthonormal}), the coefficients are given by:

\begin{equation}
\label{kappa}
\kappa _i  = \sum\limits_{n = 1}^{N} {f _i^* \left[ n \right] x\left[ n \right]} 
\end{equation}

It is desired to find a particular orthonormal set of functions such that:

\begin{equation}
\label{kappa_orthonormal}
E\left[ {\kappa _i \kappa _j^* } \right] = \left\{ \begin{array}{l}
 \varsigma _j^2 \begin{array}{*{20}c}
   {} & {}  \\
\end{array}i = j \\ 
 0_{}^{} \begin{array}{*{20}c}
   {} & {}  \\
\end{array}i \ne j \\ 
 \end{array} \right.
 \end{equation}

Therefore, the coefficients are uncorrelated. Define random vector ${\bf x} = \left[ {x\left[ 1 \right],x\left[ 2 \right], \cdots ,x\left[ N \right]} \right]^T $, coefficient vector ${\bf \kappa } = \left[ {\kappa _1 ,\kappa _2 , \cdots ,\kappa _N } \right]^T$, and the matrix

\begin{equation}
\label{Phi_Matrix}
{\bf \Phi } = \left[ {{\bf \phi }_1 ,{\bf \phi }_2 , \cdots ,{\bf \phi }_N } \right];
\end{equation}

where

\begin{equation}
\label{Phi_vector}
{\bf \phi }_i  = \left[ {f _i \left[ 1 \right],f _i \left[ 2 \right], \cdots ,f _i \left[ {N} \right]} \right]^T
\end{equation}

From (\ref{orthonormal}), ${\bf \Phi}$ is a unitary matrix such that ${\bf \Phi \Phi }^{*T}  = {\bf I}_N $. (\ref{x_expansion}) and (\ref{kappa_orthonormal}) can now be expressed in matrix formulation as:

\begin{equation}
\label{x_matrix}
{\bf x} = {\bf \Phi \kappa }
\end{equation}

and

\begin{equation}
\label{kappa_matrix}
{\bf \kappa } = {\bf \Phi }^{*T} {\bf x}
\end{equation}

Equation (\ref{x_matrix}) and (\ref{kappa_matrix}) have the following interpretation. If we consider the sequence $x\left[ n \right]$ as a vector ${\bf x}$ in an $N$-dimensional space, then $\kappa _i$ can be regarded as components of the same vector with respect to a rotated coordinate system. If we choose ${{\bf \phi }_i}$ as the eigenvectors of the covariance matrix:

\begin{equation}
\label{Cov_x}
{\bf R}_{\bf x}  = E\left[ {{\bf xx}^{*T} } \right]
\end{equation}

Then, the resulting $\kappa _i$ satisfy (\ref{kappa_orthonormal}). Therefore, the desired set of basis functions in (\ref{x_expansion}) are determined by the eigenvectors of the covariance matrix ${\bf R} _{\bf x}$:

\begin{equation}
\label{eigen_func}
{\bf R}_{\bf x} {\bf \phi }_i  = \lambda _i {\bf \phi }_i 
\end{equation}

Thus in the new coordinate system, eigenvectors ${{\bf \phi }_i}$ determine the directions, while eigenvalues $\lambda _i$ determine the signal energy in the corresponding directions.

The transformation in (\ref{kappa}) with such basis functions is the DKLT, and (\ref{x_expansion}) is called the Karhunen--Lo{\`e}ve expansion for the random process. The DKLT is the only transformation that results in (\ref{kappa_orthonormal}). 

\subsection{Properties of Feature and DKLT}

DKLT has many useful properties. They have been successfully used in principle component analysis (PCA) \cite{smith2002tutorial}, singular spectrum analysis (SSA) \cite{ghil2002advanced} and pattern recognition \cite{fukunaga1970application}, etc.. We list two of the properties for spectrum sensing. 

\textsl{Property 1:}

Equation (\ref{x_expansion}) is the optimal linear approximation representation of the random process if the expansion is truncated to use $M < N$ orthonormal basis functions:

\begin{equation}
\label{opt_expansion}
\hat x\left[ n \right] = \sum\limits_{i = 1}^M {\kappa _i f _i \left[ n \right]} ;\begin{array}{*{20}c}
   {} & {}  \\
\end{array}M < N
\end{equation}

\textsl{Property 2:}

The leading eigenvector ${\bf \phi }_1 $ is determined by the direction with largest signal energy. For any $\lambda_1 > \sigma_w^2$, ${\bf \phi }_1$ will remain almost the same. 

\begin{figure}[tbp]
	\centering
		\includegraphics[width=0.50\textwidth]{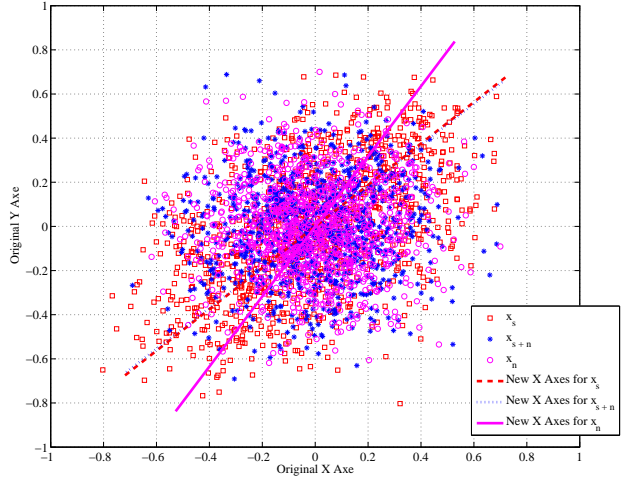}
		\caption{Illustration of \textsl{Property 2}.}
	\label{fig:New_Directions_of_Signals}
\end{figure}

\textsl{Property 2} has a geometric explanation with a two dimensional case in Fig. \ref{fig:New_Directions_of_Signals}. Assume we have $2 \times 1$ random vectors $x_{s+n} = x_s + x_n$, where $x_s$ is vectorized sine sequence and $x_n$ is the vectorized WGN sequence. SNR is set to $0$ dB. There are 1000 samples for each random vectors in Fig. \ref{fig:New_Directions_of_Signals}. Now we use DKLT to set the new $X$ axes for each random vector samples such that $\lambda_1$ is strongest along the corresponding new $X$ axes. It can be seen that new $X$ axes for $x_s$ (SNR = $\infty$ dB) and $x_{s+n}$ (SNR = $0$ dB) are almost the same. $X$ axes for $x_n$ (SNR = $-\infty$), however, is rotated with some random angle. This is because WGN has almost same energy distributed in every direction. New $X$ axes for noise will be random and unpredictable but the direction for signal is very robust, as long as $\lambda_1 > \sigma_w^2$. 

\textsl{Property 2} helps us to conclude that among all eigenvectors, only leading eigenvector is most robust against noise. Together with \textsl{Property 1}, we can prove that the leading eigenvector is also optimal approximation of original signal by simply setting $M$ to $1$ in (\ref{opt_expansion}). Moreover, since signal energy/noise energy estimation is not used in the entire process \cite{kay1998fundamentals}, there is no noise uncertainty problem for feature learning.

In another view, the effective SNR on the leading eigenvector is higher than the original SNR. If we do DKLT on $s\left[ n \right]$ and $w\left[ n \right]$, we have eigenvalues $\lambda _{s,i}$ for $s\left[ n \right]$ and $\lambda _{w,i} $ for $w\left[ n \right]$; eigenvectors ${\bf \phi }_{s,i}$ for $s\left[ n \right]$ and ${\bf \phi }_{w,i}$ for $w\left[ n \right]$. The SNR for $x\left[ n \right]$ is:

\begin{equation}
\label{SNR_x}
{\rm SNR}_x = \sum\limits_{i = 1}^N {\lambda _{s,i} } /\sum\limits_{i = 1}^N {\lambda _{w,i} }
\end{equation}

Suppose we only use the leading eigenvector ${\bf \phi }_1 $ to approximate $x\left[ n \right]$ by $\hat x\left[ n \right]$ in (\ref{opt_expansion}). Since $w \left[ n \right]$ is white, $\lambda _{w,i} = \sigma _w^2 $ for all $i$, the SNR for $\hat x\left[ n \right]$ is therefore 

\begin{equation}
\label{SNR_hat}
{\rm SNR}_{\hat x} = \lambda _{s,1} /\sigma _w^2
\end{equation}

The SNR gain after using the leading eigenvector of DKLT is:

\begin{equation}
\label{ratio}
G_{{\rm SNR}}  = \frac{{{\rm SNR}_{\hat x} }}{{{\rm SNR}_x }} = \frac{{N\lambda _{s,1} }}{{\sum\limits_{i = 1}^N {\lambda _{s,i} } }}
\end{equation}

Since $\lambda _{s,1}  \ge \lambda _{s,2}  \ge  \cdots \lambda _{s,N} $, $G_{{\rm SNR}} \ge 1$. Such SNR gain is optimal \cite{makhoul1981eigenvectors}. 

This is the foundation of the FLA and FTM. They use the fact that consecutive features of WSS signal are similar, while consecutive features of noise are random. 

%
%
%
%
%

\subsection{Implementation Issue}
The major processing part of our algorithms lies in the feature extraction. If we analyze the processing delay of the feature extraction, it can be divided to two steps. First is to compute the covariance matrix ${\bf \tilde R}_{x,m}$ in (\ref{Rx_segment}) and the other is the eigenvector calculation. Since the computation of ${\bf \tilde R}_{x,m}$ can be done in real-time, the major processing delay lies in the eigenvector calculation. In the feature extraction, we only want to know the leading eigenvector and we do not need to do the complete eigen-decomposition, which has computation complexity of ${\rm O}\left( {N^3 } \right)$. Recently, a fast PCA algorithm for fixed point implementation have been proposed with computation complexity ${\rm O}\left( {N^2 } \right)$ \cite{sharma2007fast}, minimizing the processing delay. Being able to compute (\ref{Rx_segment}) in real-time is a huge advantage if compared with spectral methods using the fast Fourier transform (FFT). Since FFT can only be calculated when all $N _s$ samples are captured, the computation complexity is ${\rm O}\left( {N_s \log \left( {N_s } \right)} \right)$. Because usually $N_s >> N$, the feature extraction has much less delay than FFT. Currently we have implemented the algorithms in field programmable gate array (FPGA) and digital signal processor (DSP) \cite{qiu2010towards,zhang_2010_letter}.

\section{Simulation Results}
In this section, we first use show that the signal feature can be extracted under unknown low SNR. Then, real-world captured data is used to show that the signal feature can be learned blindly and is stable over time. Finally, we compare the detection performance of the FTM with the MME in very low SNR, using the same real-world data. In the comparison, same covariance matrix is used, but the FTM has the learned feature as prior knowledge.

\subsection{Feature Robustness Test Against Noise}
\begin{figure}[tbp]
	\centering
		\includegraphics[width=0.50\textwidth]{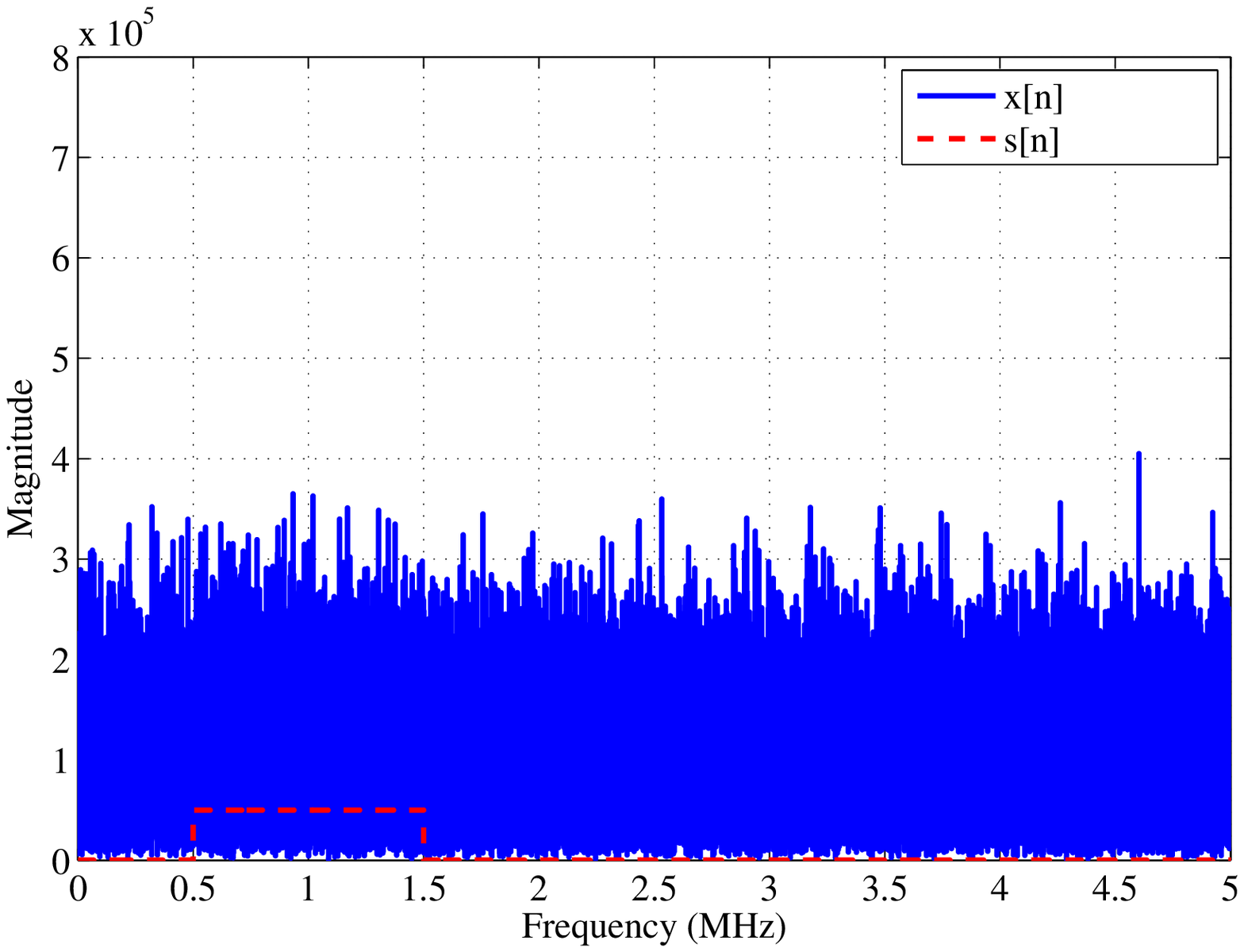}
	\caption{Spectrum of $x \left[ n \right]$ and $s \left[ n \right]$.}
	\label{fig:Wideband_DFT}
\end{figure}

\begin{figure}[tbp]
	\centering
		\includegraphics[width=0.50\textwidth]{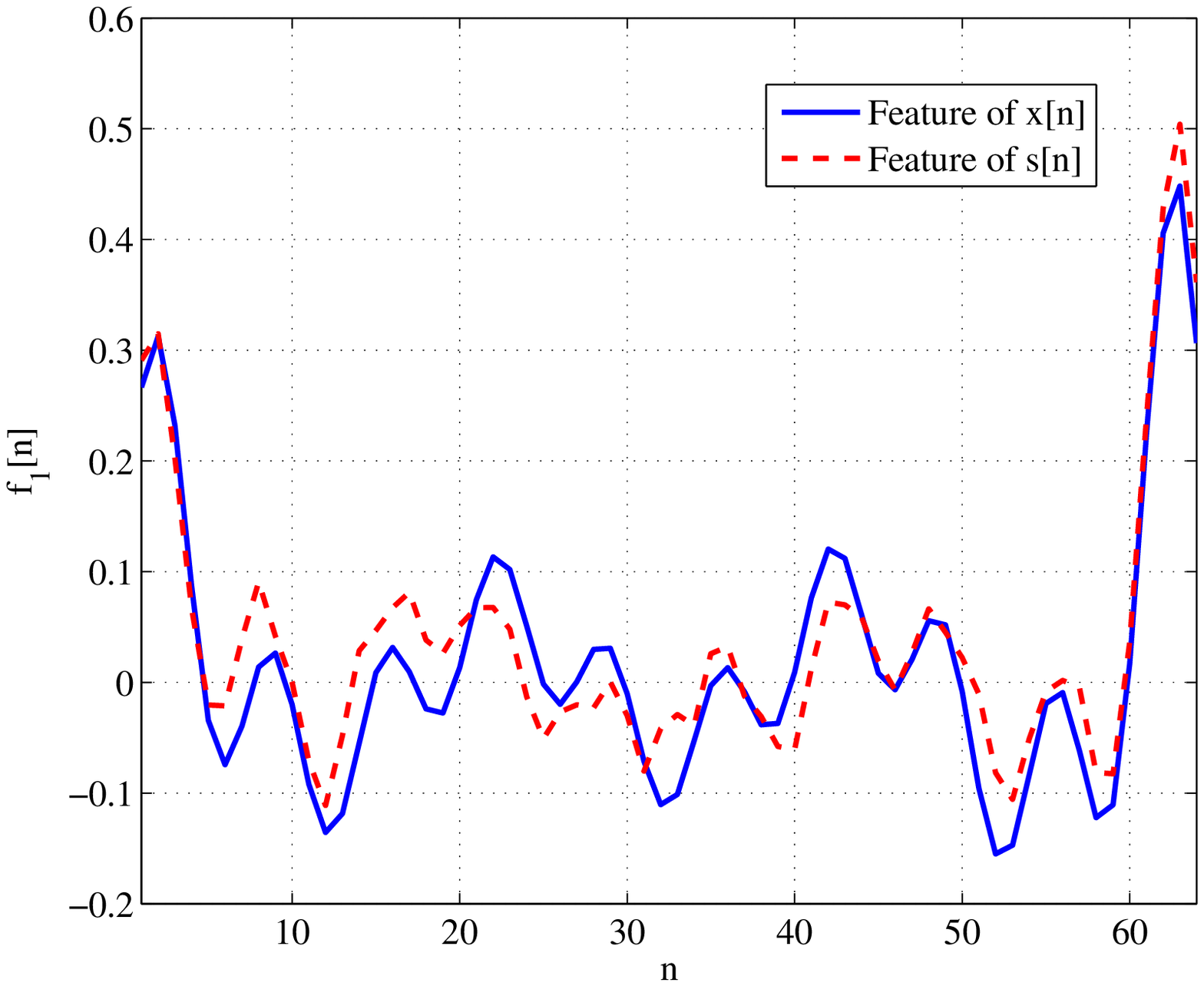}
	\caption{Features of $x \left[ n \right]$ and $s \left[ n \right]$.}
	\label{fig:Wideband_Feature}
\end{figure}

Here we give an example to show feature extraction under unknown low SNR. $s \left[ n \right]$ has constant power spectrum density in the frequency band from $0.5$ MHz to $1.5$ MHz. $x \left[ n \right]$ is the noisy signal with unknown amount of noise. As shown in Fig. \ref{fig:Wideband_DFT}, no spectral information of $x \left[ n \right]$ can be extracted in the frequency domain. We use the FLA to extract the features from the noisy $x \left[n \right]$ and the noise free $s \left[n \right]$. As can be seen in Fig. \ref{fig:Wideband_Feature}, those two features are very similar, with similarity as high as $94\%$. As a result, feature is very robust against noise and there is no noise uncertainty in feature extraction.

\subsection{Feature Learning Test with Real-world Data}
\begin{figure}[tbp]
	\centering
		\includegraphics[width=0.50\textwidth]{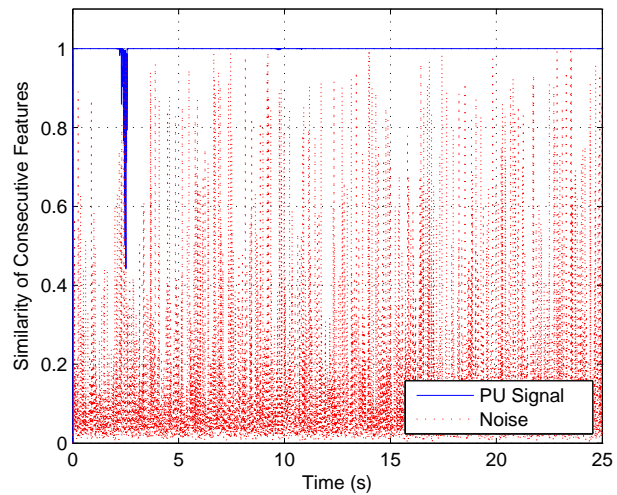}
	\caption{Similarity of consecutive features of PU signal and noise in 25 seconds.}
	\label{fig:Fading_Feat_Corr}
\end{figure}

Then we use real world data to demonstrate that the signal feature is very stable over time while the noise feature is random. Field measurements of DTV done in Washington D.C. \cite{DTV2006Measurements} are used as the PU signal. Simulated WGN samples are used. The captured signal has a duration of about 25 seconds. All synchronization information of the DTV signal is blind to the SU receiver. Receiver SNR and the communication channel between the transmitter and receiver are also unknown. However, we do know that the received SNR is changing at the receiver and the channel has slow fading. We use the FLA to calculate the similarities of consecutive features for both signal and noise in 25 seconds, respectively. We set $N_s = 10^5$ and $N = 64$. The corresponding duration of each segment is approximately $4.6$ ms. Fig. \ref{fig:Fading_Feat_Corr} shows Similarity of Consecutive Features VS Time plot. By setting $T_e = 90\%$, $\rho_{i, i+1} > T_e$ for $99.46\%$ amount of time when the PU signal exists. Moreover, the similarity between the features of the first sensing segment and the last sensing segment is as high as $99.98\%$, showing that the signal feature is very stable and almost unchanged in 25 seconds. When the PU signal does not exist, $\rho_{i, i+1} > T_e$ for only $0.92\%$ amount of time. Sohpisticated learning algorithms will be developed in the feature learning process to obtain the signal feature in a robust and fast manner.

\subsection{ROC Curves for the FTM and the MME}
\begin{figure}[tbp]
	\centering
		\includegraphics[width=0.50\textwidth]{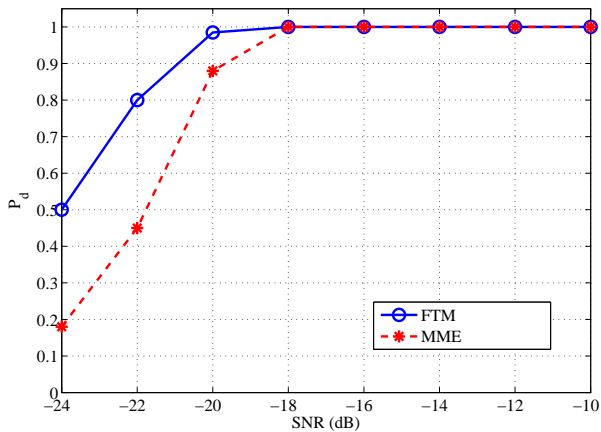}
	\caption{$P_d$ VS SNR for FTM and MME. $N_s = 10^5$, $N = 64$}
	\label{fig:FD_VS_MME_SNR_Curve}
\end{figure}

\begin{figure}[tbp]
	\centering
		\includegraphics[width=0.50\textwidth]{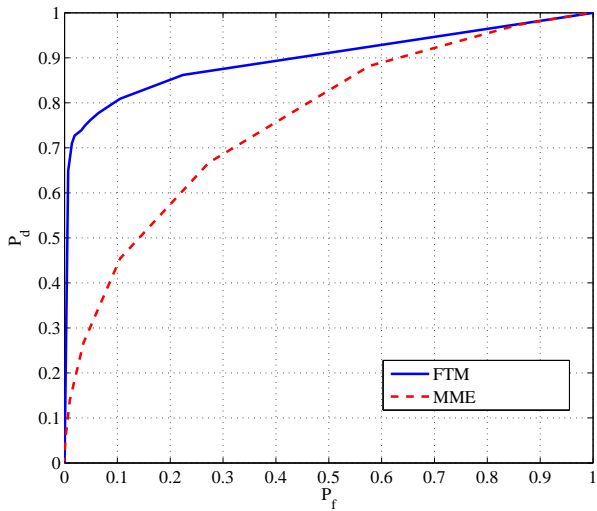}
	\caption{ROC of FTM and MME. SNR = -22 dB,  $N_s = 10^5$, $N = 64$}
	\label{fig:ROC_-22dB_64}
\end{figure}

We use one segment of the previous DTV data samples as the clean received PU signal $s\left[ n \right]$ and add noise with variance $\sigma _w^2 $ to emulate $w\left[ n \right]$. In the simulation, the signal feature is the prior knowledge. Sensing time is set to approximately $4.6$ ms with $N _s = 10^5$ and $N = 64$. WGN is added according to different SNR levels. We compare the results of both algorithms. The MME uses the covariance matrix's max-min eigenvalue ratio, $\lambda_{\rm max} / \lambda_{\rm min}$ for detection \cite{zeng2007maximum}, and uses no prior knowledge. $N _s$ and $N$ are set the same for these two algorithms. In the simulation, we perform both algorithms on the same signal and noise and repeat the simulation for $1000$ times. Fig. \ref{fig:FD_VS_MME_SNR_Curve} shows the $P_d$ VS SNR, with $P _f = 10 \%$. It can be seen that to reach $P_d \approx 100\%$, the minimum required SNR for the FTM is about $2$ dB lower than that of the MME. Note that in our simulations, all $T_f$ set by the FTM to get $P_f = 10\%$ are very stable for different SNR. This is because $T_f$ is independent of SNR, signal energy or noise energy, and the FTM does not have noise uncertainty problem. Fig. \ref{fig:ROC_-22dB_64} shows the receiver operating characteristic (ROC) curves when SNR = $-22$ dB. At $P_f = 10\%$, FTM has $P_d = 80\%$, while MME only has $P_d = 42\%$. This shows the advantage of using the prior knowledge.

\section{Conclusions}
Signal feature is location dependent. We propose to learn signal feature blindly and use it for spectrum sensing. We define the signal feature as the leading eigenvector of signal's sample covariance matrix, because it is a robust approximation of original signal against noise and optimum in effective SNR, based on DKLT properties. We propose the FLA for blind feature learning and the FTM for spectrum sensing with the signal feature as the prior knowledge. Since our algorithms do not depend on SNR or the noise energy, noise uncertainty problem is successfully avoided. We use simulated data and real-world data to demonstrate feature's robustness against noise and its stability over time. Detection performance of the FTM in low SNR is compared with MME, which is totally blind. Simulation results show that to achieve $P_d \approx 100\%$ and $P_f = 10\%$, the minimum requried SNR for the FTM is about $2$ dB lower than that of the MME.

This is only the beginning of our work. Further research topics include implementation, sophisticated feature learning and quantizing the thresholds, etc. In addition, feature extracted by DKLT is optimum only in the context of linear transforms. When signal has non-linear structures, non-linear methods like Kernel-PCA \cite{scholkopf1999kernel} and manifold-learning \cite{weinberger2006unsupervised} can be the next powerful tools to be explored.

\label{conclusions}

\section*{Acknowledgment}
This work is funded by National Science Foundation through grants (ECCS-0901420),  (ECCS-0821658), and Office of Naval Research through two contracts (N00014-07-1-0529, N00014-11-1-0006).

\bibliographystyle{ieeetr}
\bibliography{bib/CR,bib/Pattern_Recognition,bib/CR_Peng}

\begin{thebibliography}{10}

\bibitem{staple2004end}
G.~Staple and K.~Werbach, ``{The end of spectrum scarcity},'' {\em IEEE
  Spectrum}, vol.~41, no.~3, pp.~48--52, 2004.

\bibitem{force2002spectrum}
FCC, ``{Spectrum policy task force report},'' tech. rep., ET Docket No. 02-155,
  Nov. 2002.

\bibitem{cabric2004implementation}
D.~Cabric, S.~Mishra, and R.~Brodersen, ``{Implementation issues in spectrum
  sensing for cognitive radios},'' in {\em Asilomar Conference on Signals,
  Systems, and Computers}, vol.~1, pp.~772--776, 2004.

\bibitem{ieee80222}
``Ieee 802.22 working group on wireless regional area networks,'' 2004.
\newblock http://www.ieee802.org/22.

\bibitem{cordeiro2007spectrum}
C.~Cordeiro, M.~Ghosh, D.~Cavalcanti, and K.~Challapali, ``Spectrum sensing for
  dynamic spectrum access of {TV} bands,'' in {\em Second International
  Conference on Cognitive Radio Oriented Wireless Networks and Communications},
  (Orlando, Florida), Aug. 2007.

\bibitem{Quan2009ss}
Z.~{Quan}, S.~J. {Shellhammer}, W.~{Zhang}, and A.~H. {Sayed}, ``Spectrum
  sensing by cognitive radios at very low {SNR},'' in {\em IEEE Global
  Communications Conference 2009}, 2009.

\bibitem{dandawate1994statistical}
A.~Dandawate and G.~Giannakis, ``{Statistical tests for presence of
  cyclostationarity},'' {\em IEEE Transactions on Signal Processing}, vol.~42,
  no.~9, pp.~2355--2369, 1994.

\bibitem{DTV2006Measurements}
V.~Tawil, ``51 captured {DTV} signal.'' {http://grouper.ieee.org/groups
  /802/22/Meeting\_documents/2006\_May/Informal\_Documents}, May 2006.

\bibitem{tandra2005fundamental}
R.~Tandra and A.~Sahai, ``{Fundamental limits on detection in low SNR under
  noise uncertainty},'' in {\em Proc. Wireless Commun. Symp. on Signal
  Process.}, Jun. 2005.

\bibitem{zeng2007maximum}
Y.~Zeng and Y.~Liang, ``{Maximum-minimum eigenvalue detection for cognitive
  radio},'' in {\em IEEE 18th International Symposium on Personal, Indoor and
  Mobile Radio Communications (PIMRC) 2007}, pp.~1--5, 2007.

\bibitem{lim2008glrt}
T.~Lim, R.~Zhang, Y.~Liang, and Y.~Zeng, ``{GLRT-based} spectrum sensing for
  cognitive radio,'' in {\em Global Telecommunications Conference}, pp.~1--5,
  IEEE, 2008.

\bibitem{watanabe1965kle}
S.~Watanabe, ``Karhunen--{L}o{\`e}ve expansion and factor analysis:
  theoreticalremarks and applications,'' in {\em Proc. 4th Prague (IT) Conf.},
  pp.~635--660, 1965.

\bibitem{young1971reliability}
T.~Y. Young, ``{The reliability of linear feature extractors},'' {\em IEEE
  Transactions on Computers}, vol.~C-20, no.~9, pp.~967--971, 1971.

\bibitem{therrien1992discrete}
C.~Therrien, {\em {Discrete Random Signals and Statistical Signal Processing}}.
\newblock Englewood Cliffs, NJ: Prentice Hall PTR, 1992.

\bibitem{smith2002tutorial}
L.~Smith, ``{A tutorial on principal components analysis},'' {\em Cornell
  University, USA}, vol.~51, p.~52, 2002.

\bibitem{ghil2002advanced}
M.~Ghil, M.~Allen, M.~Dettinger, K.~Ide, D.~Kondrashov, M.~Mann, A.~Robertson,
  A.~Saunders, Y.~Tian, F.~Varadi, {\em et~al.}, ``{Advanced spectral methods
  for climatic time series},'' {\em Rev. Geophys}, vol.~40, no.~1, p.~1003,
  2002.

\bibitem{fukunaga1970application}
K.~Fukunaga and W.~Koontz, ``Application of the {K}arhunen--{L}o{\`e}ve
  expansion to feature selection and ordering,'' {\em IEEE Transactions on
  Computers}, vol.~C-19, no.~4, pp.~311--318, 1970.

\bibitem{kay1998fundamentals}
S.~Kay, {\em Fundamentals of Statistical Signal Processing, Volume 2: Detection
  theory}.
\newblock Prentice Hall PTR, 1998.

\bibitem{makhoul1981eigenvectors}
J.~Makhoul, ``{On the eigenvectors of symmetric Toeplitz matrices},'' {\em IEEE
  Transactions on Acoustics, Speech and Signal Processing}, vol.~29, no.~4,
  pp.~868--872, 1981.

\bibitem{sharma2007fast}
A.~Sharma and K.~Paliwal, ``{Fast principal component analysis using
  fixed-point algorithm},'' {\em Pattern Recognition Letters}, vol.~28, no.~10,
  pp.~1151--1155, 2007.

\bibitem{qiu2010towards}
R.~Qiu, Z.~Chen, N.~Guo, Y.~Song, P.~Zhang, H.~Li, and L.~Lai, ``Towards a
  real-time cognitive radio network testbed: Architecture, hardware platform,
  and application to smart grid,'' in {\em Networking Technologies for Software
  Defined Radio (SDR) Networks, 2010 Fifth IEEE Workshop on}, pp.~1--6, IEEE,
  2010.

\bibitem{zhang_2010_letter}
P.~Zhang, R.~Qiu, and G.~Nan, ``Demonstration of feature learning based
  spectrum sensing in cognitive radio.'' Submitted, 2010.

\bibitem{scholkopf1999kernel}
B.~Sch{\"o}lkopf, A.~Smola, and K.~M{\"u}ller, {\em {Advances in Kernel
  Methods--Support Vector Learning}}.
\newblock Cambridge, MA: MIT Press, 1999.

\bibitem{weinberger2006unsupervised}
K.~Weinberger and L.~Saul, ``{Unsupervised learning of image manifolds by
  semidefinite programming},'' {\em International Journal of Computer Vision},
  vol.~70, no.~1, pp.~77--90, 2006.

\end{thebibliography}

\end{document}